# TRANSITION WAKE-FIELDS IN RESISTIVE TUBE

M. Ivanyan, V. Tsakanov

Center for the Advancement of Natural Discoveries using Light Emission - CANDLE

Acharyan 31, Yerevan, 375040, Armenia

A. Tsakanian

Yerevan State University, Alek Manukyan 1, Yerevan, 375025, Armenia

*Abstract*

The transition wake fields excited by relativistic point charge in cylindrical resistive round pipe are studied. The problem has been solved for the infinite pipe with abrupt change of the walls conductivity from perfect boundary condition to resistive. The analytical presentation of the longitudinal monopole wake field is given.

## INTRODUCTION

The development of the physics of ultra-short, ultra-low emittance relativistic electron beams drives to the necessity of exact derivation of the wake fields induced by point charge during the interaction with its environment. Although large number of such electro-dynamical problems in steady state regime have been well understood and derived in the terms of the structure impedance and wake potentials [1,2], the transition effects in some cases are still missed from the general picture. The importance of the knowledge on the transition behavior of the excited fields is driven also by both better understanding of physical processes and the proper application of the numerical codes to evaluate the excited fields in real structure.

The radiation excited by the relativistic charged particle in the infinite round pipe with finite conductivity wall material is presented in number of Refs. [1-6]. The impedance of finite length resistive tube has been derived in Refs. [7,8].

In this paper, the study of the longitudinal transition wake fields in resistive tube is given. The explicit presentation for the transient wake fields is given when the charge travels along the axis of cylindrically symmetric resistive pipe. It is shown that the characteristic distance $s_0$ [1,5] used in resistive wake field derivation defines the typical transition length of steady state wake field formation.

In our calculations we will drop a very high frequency term arising from the displacement current, a term which becomes important only when $\omega \sim \omega_p$, with $\omega_p$ the plasma frequency of the free electrons in the metal. Although this term is the source of the secondary fields, the energy loss is negligibly small [9] and is usually neglected in the classical treatment of the problem [6].

## TRANSIENT WAKE FIELDS AND IMPEDANCE

Consider the ultra-relativistic point charge $q$ moving with the velocity of light $c$ along the $z$-axis of infinite round pipe with the abrupt change of the walls conductivity at position $z = 0$ (Fig.1). The pipe walls are the perfectly conducting material at $z < 0$ and have the finite conductivity at $z > 0$, the pipe radius $b$ is constant and the wall thickness is infinite. In perfectly conducting pipe the charge with the constant velocity does not excite electromagnetic fields as the image charge on wall surface is propagating synchronously with charge inside the vacuum pipe. Starting from the transition at the point $z = 0$ the surface current is retarding and the charge field lines are perturbed to satisfy new boundary condition. The electromagnetic field of the charge is penetrating to the metallic walls leading to the Ohmic losses in the walls. The energy is taken from the energy of the charge and the charge excites the wake fields. The transition behavior extends unless the retarding surface current reaches its steady state regime, in the meantime the steady state wake fields associated with the charge are formed.

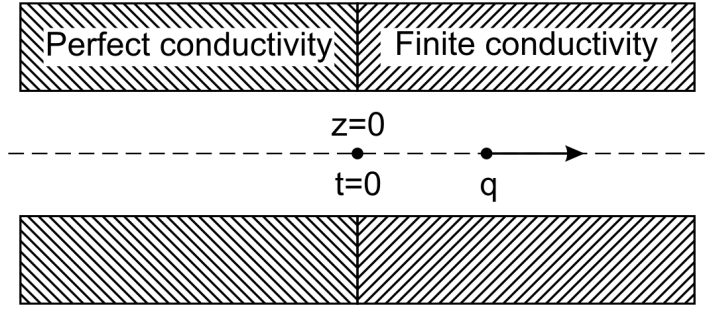

Fig.1. The geometry of the problem

The boundary conditions on the inner surface of the perfectly conducting walls $(z < 0, r = b)$ are given by the vanishing of the tangential electrical component of the field. In the pipe region with finite conductivity $(z > 0, r = b)$ we assume the Leontovich boundary conditions [10] that are valid for metallic surface.

Let us start with relativistic charge $q$ moving with the velocity of light $c$ in perfectly conducting pipe. The nonzero Fourier components of the electromagnetic fields synchronously moving with the charge are given by

$$E_r = \frac{q}{2\pi r \varepsilon_0 c} e^{-ikz} \qquad H_\theta = \frac{q}{2\pi r} e^{-ikz} \tag{1}$$

where $\omega$ is the frequency, $\varepsilon_0$ is the dielectric constant, $k = \omega/c$ is the wave number. In resistive pipe ($z > 0$) the nonzero Fourier components of the fields satisfy the following equation

$$\frac{1}{r}\frac{\partial}{\partial r} r \frac{\partial}{\partial r} E_z + \frac{\partial^2 E_z}{\partial z^2} - k^2 E_z = 0$$
$$\frac{1}{r}\frac{\partial r E_r}{\partial r} = -\frac{\partial E_z}{\partial z} + 4\pi\rho \tag{2}$$
$$\frac{\partial B_\theta}{\partial z} = -ikE_r$$

with $\rho = \frac{q}{r}\delta(r)e^{-ikz}$ and the Leontovich boundary condition at $r = b$

$$E_z = -\zeta \cdot B_\theta \tag{3}$$

with $\zeta = \sqrt{j\mu_0\omega/\sigma_c}$ metallic surface complex impedance, $\sigma_c$ metal conductivity, $\mu_0$ magnetic permeability. The common solution of the homogeneous Maxwell equations finite in the structure axis ($r = 0$) is given by the Bessel function and is read as

$$E_z = uJ_0\left(u\frac{r}{b}\right)\left(Ae^{-ipz} + Be^{ipz}\right)$$
$$E_r = ibpJ_0'\left(u\frac{r}{b}\right)\left(Ae^{-ipz} - Be^{ipz}\right) \tag{4}$$
$$H_\theta = -\frac{jk^2 b}{\mu\omega}J_0'\left(u\frac{r}{b}\right)\left(Ae^{-ipz} + Be^{ipz}\right)$$

with $p(\omega) = \sqrt{k^2 - [u(\omega)/b]^2}$ the propagation constant and $A(\omega), B(\omega)$ arbitrary coefficients. The boundary condition (3) along with the expression for the propagation constant defines the dispersion relation for the electromagnetic waves in resistive pipe

$$uJ_0(u) = \frac{jk^2 b\zeta}{\mu\omega} J_0'(u) \tag{5}$$

The general solution of the inhomogeneous Maxwell equations is then given by

$$
\begin{aligned}
E_z &= uJ_0\left(u\frac{r}{b}\right)\left(Ae^{-ipz} + Be^{ipz}\right) \\
E_r &= \frac{q}{2\pi r\varepsilon_0 c} e^{-ikz} - jbpJ_0'\left(u\frac{r}{b}\right)\left(Ae^{-ipz} - Be^{ipz}\right), \\
H_\theta &= \frac{q}{2\pi r} e^{-ikz} - \frac{jk^2 b}{\mu\omega} J_0'\left(u\frac{r}{b}\right)\left(Ae^{-ipz} + Be^{ipz}\right)
\end{aligned}
\tag{6}
$$

The Leontovich boundary condition for the full field is read as

$$\zeta^{-1} uJ_0(u)\left(Ae^{-jpz} + Be^{jpz}\right) = -\frac{q}{2\pi b} e^{-ikz} + \frac{jk^2 b}{\mu\omega} J_0'(u)\left(Ae^{-jpz} + Be^{jpz}\right) \tag{7}$$

that leads to $p(\omega) = k$ ($u = 0$ only the synchronous component of the field contributes into the interaction) and the following relation for the coefficient $A$

$$(uA)_{u=0} = -qZ^{st}(\omega) = -\frac{q}{2\pi b}\left[\zeta^{-1} + \frac{jk^2 b}{2\mu\omega}\right]^{-1} \tag{8}$$

The quantity $Z^{st}(\omega)$ coincides with the monopole longitudinal impedance per unit length of infinite resistive tube [1,5] and can be presented as

$$Z^{st}(\omega) = \frac{Z_0 s_0}{2\pi b^2}\left(\frac{1-i}{\sqrt{k_0}} + i\frac{k_0}{2}\right) \tag{9}$$

with $Z_0 = 120\pi$ the impedance of the free space, $k_0 = ks_0$ the dimensionless wave number and $s_0$ the characteristic distance given by

$$s_0 = \left(\frac{2cb^2\varepsilon_0}{\sigma_c}\right)^{1/3} \tag{10}$$

. The unknown coefficient $B$ is defined by the matching of the components $E_z, B_\theta$ at $z = 0$ that leads to $B = -A$. The radial component of electric field has discontinuity at $z = 0$ due to abrupt change of the boundary conditions in our model. Note, that in resistive pipe the radial component of the steady state solution has discontinuity at the metallic surface $r = b$ [2].

Thus the non-vanishing field components in resistive tube are given by

$$
\begin{aligned}
E_z &= qZ^{st}(\omega)\left(e^{jkz} - e^{-jkz}\right) \\
E_r &= \frac{q}{2\pi r\varepsilon_0 c} e^{-ikz} - \frac{jkr}{2} Z^{st}(\omega)\left(e^{jkz} + e^{-jkz}\right) \\
H_\theta &= \frac{q}{2\pi r} e^{-ikz} - \frac{jkr}{2Z_0} Z^{st}(\omega)\left(e^{jkz} - e^{-jkz}\right)
\end{aligned}
\tag{11}
$$

The corresponding longitudinal wake field in time domain induced by point charge in resistive pipe is given by the inverse Fourier transform

$$E_z = \int_{-\infty}^{\infty} Z^{st}(\omega)\left(e^{jkz} - e^{-jkz}\right)e^{j\omega t}d\omega \qquad (12)$$

The impedance $Z^{st}(\omega)$ has two resudes $k_0 = \pm\sqrt{3} + i$ and the second term of (12) corresponds to the longitudinal wake potential of the point charge in infinitely long pipe [5,6].

$$E_z^{st}(s) = -\int_{-\infty}^{\infty} Z^{st}(\omega)e^{j(ct-z)\frac{\omega}{c}}d\omega = -\frac{4}{\pi\varepsilon_0 b^2}\left[\frac{1}{3}e^{-s/s_0}\cos\left(\sqrt{3}\frac{s}{s_0}\right) - \frac{\sqrt{2}}{\pi}\int_0^{\infty}dx\frac{x^2}{x^6+8}e^{-x^2\frac{s}{s_0}}\right] \qquad (13)$$

where $s = ct - z$ is the distance behind the driving charge. Note, that the field vanish ($E_z^{st}(s) = 0$) in front of the charge ($s < 0$) due to causality principle. For our geometry this term represents the steady state solution.

The first term in (12) can be evaluated similarly to steady state solution

$$E_z^{tr}(z,t) = \int_{-\infty}^{\infty} Z^{st}(\omega)e^{j(ct+z)\frac{\omega}{c}}d\omega = E_z^{st}(z+ct) \qquad (14)$$

Thus the longitudinal wake field excited in the resistive pipe ($z > 0$) is given by

$$E_z(r,z,t) = E_z^{st}(ct-z) - E_z^{st}(ct+z) \qquad (15)$$

with the causality principle of vanishing electromagnetic fields $E_z(z,t) = 0$ in front of the charge ($z > ct$) as for $t = 0$ the longitudinal component of electromagnetic field is zero for $z > 0$. Note, that for any $t > 0$ the electromagnetic field components $E_z, B_\theta$ vanish at $z = 0$.

The monopole longitudinal impedance $Z(\omega)$ per unit length in resistive part of the tube is given by

$$Z(\omega) = \frac{1}{L}\int_0^L E_z(\omega,z)e^{ikz}dz \qquad (16)$$

and according to (12) is presented by the sum of the steady-state impedance $Z^{st}(\omega)$ and the transition impedance $Z^{tr}(\omega)$ caused by the abrupt change of the wall conductivity at $z = 0$

$$Z(\omega) = Z^{st}(\omega) + Z^{tr}(\omega) \qquad (17)$$

with

$$Z^{tr}(\omega) = Z^{st}(\omega)\frac{e^{2jkL} - 1}{2jkL} \qquad (18)$$

where $L$ is the length of resistive part of the tube. For the large parameter $kL$ ($kL \gg 1$) the contribution of the transition impedance to general impedance vanishes.

## DISSCUSSIONS

Excited wake fields in the resistive pipe are the superposition of the steady state wake-fields and the transition fields. While the steady state electromagnetic fields depend on the distance behind the driving charge $s = ct - z$, the transition fields depend on both the driving charge position $z$ and time $t$. The excited fields are zero for $t < 0$ (before the charge reaches the transition) and for $s < 0$ (in front of the charge). Fig. 2 presents the dependence of normalized retarding electric field experienced by the charge $U_0 = -E_z/E_z^{st}$ ($s = 0$) from the charge travel distance $z_q$ in resistive tube ($z_q \geq 0$). The field is normalized to the retarding electric field at $s = 0$ of steady state solution. The charge response to tube surface conductivity change is essential at the charge travel distance of $3s_0$ when the retarding field smoothly transforms to steady state regime. Thus the physical meaning of the distance $s_0$ actually is the characteristic transition length of steady state wake field formation in resistive tube.

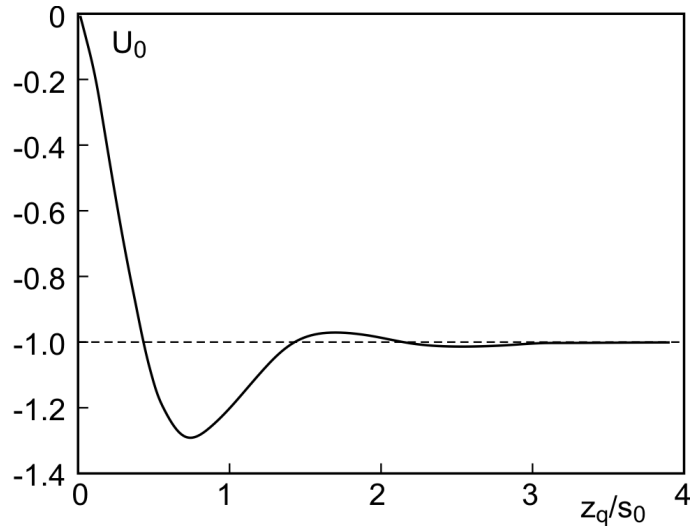

Fig. 2 Retarding electric field seen by driving charge after passing the transition at $z = 0$.

Fig. 3 presents the normalized longitudinal wake fields $U(s) = -E_z(s)/E_z^{st}(0)$ for various traveling distances $z_q$ of the driving charge after its passage the transition at $z = 0$. Again the steady state solution reached at the driving charge travel distance of $z_q = 3s_0$.

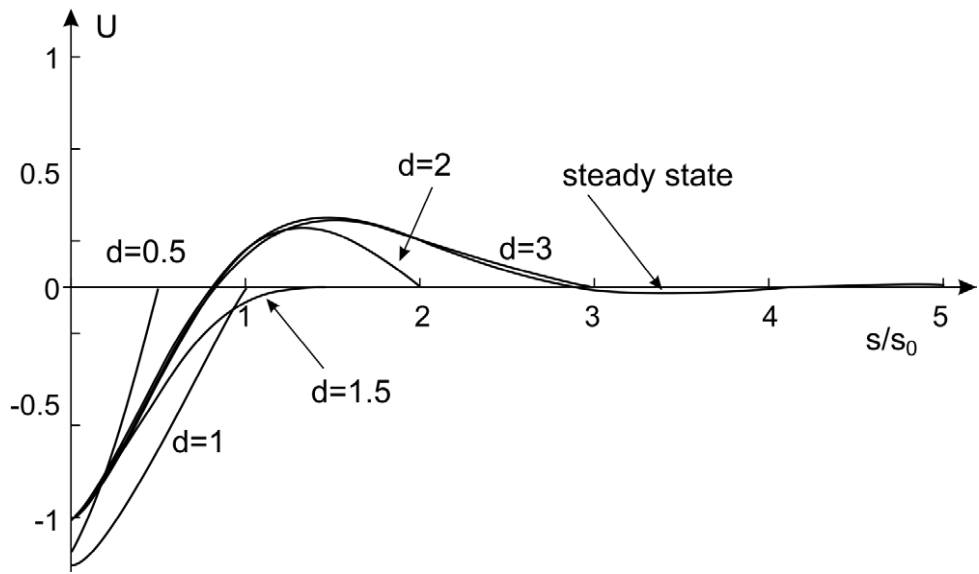

Fig.3 The longitudinal wake field behind the driving charge after passing the transition at $z = 0$. Charge position in resistive tube is given by $d = z_q / s_0$ ($d = 0.5, 1, 1.5, 2, 3$).

**SUMMARY**

We examined the transition behavior of the wake fields excited by ultrarelativistic charge in resistive pipe. The analytical presentation of the excited longitudinal electric field is given, that shows the smooth transition to the steady state regime at charge travel distance of few $s_0$ - the characteristic distance in resistive wakefield derivation. The impedance per unit length is derived in transition regime.

The results of the work can be useful for evaluation of the transition wakefields and impedances in the vacuum chamber with non-uniform metallic boundary conditions along the charge trajectory.

**REFERENCES**

1. A.W. Chao, *Physics of Collective Beam Instabilities in High Energy Accelerators* (Wiley, New York, 1993).
2. B.W.Zotter and S.A. Kheifetz, *Impedances and Wakes in High-Energy Particle Accelerators* (World Scientific, Singapore, 1997).
3. A Piwinski, Report No DESY HERA 92-11, 1992, p. 19.
4. O. Henry and O. Napoly, Part. Accel. **35**, 235 (1991).
5. K. Bane, Report No SLAC-AP-87, 1991. p.26.
6. K.L.F. Bane and M Sands, Report No SLAC-PUB-95-7074, 1995, p. 19.
7. S. Krinsky, B. Podobedov and R.L.Glukstern, Phys. Rew. ST Accel. Beams $\underline{7}$, 111401 (2004).
8. G. Stupakov, Phys. Rew. ST Accel. Beams $\underline{8}$, 044401 (2005).
9. J.D. Jackson, *Classical Electrodynamics*, (Wiley, New York, 1962).
10. L. D. Landau and E. Lifshitz, *Electrodynamics of Continuous Media*, Course of Theoretical Physics, Vol. 8 (Pergamon, London, 1960), 2nd ed.